\documentclass[12pt]{article}  
\usepackage{cite}

\normalbaselines
\catcode `\@=11
\@addtoreset{equation}{section}

\def\section{\@startsection {section}{1}{\z@}{-3.5ex plus -1ex minus
     -.2ex}{2.3ex plus .2ex}{\normalsize\bf}}
\def\subsection{\@startsection{subsection}{2}{\z@}{-3.25ex plus -1ex minus
 -.2ex}{1.5ex plus .2ex}{\normalsize\bf}}

\def\thebibliography#1{\section*{References\markboth
  {REFERENCES}{REFERENCES}}\list
  {[\arabic{enumi}]}{\settowidth\labelwidth{[#1]}\leftmargin\labelwidth
  \advance\leftmargin\labelsep
  \usecounter{enumi}}
  \def\newblock{\hskip .11em plus .33em minus -.07em}
  \sloppy
  \sfcode`\.=1000\relax}
 
\catcode `\@=12
\begin{document}
 
\vspace*{1.0cm} 
        \begin{center}
\noindent

{ \bf 
QUANTIZATION OF GRAVITY: YET ANOTHER WAY\footnote{
Partially based on author's talks 
at the XV Workshop on Geometrical Methods in Physics 
(Bia\l owie\. za, Poland, 1998) 
and XXXI Symposium on Mathematical Physics (Toru\'n, Poland, 1999). 
To appear in {\em Rep. Math. Phys. } (2000). }  
}\vspace{1.3cm}\\
        \end{center}
\noindent
\hspace*{1in}
\begin{minipage}{13cm}
Igor V. Kanatchikov$^{a,b}$  \vspace{0.3cm}\\
 $^{a}$ Laboratory of Analytical Mechanics and Field Theory \\
\makebox[3mm]{ }Institute of Fundamental Technological Research \\
\makebox[3mm]{ }Polish Academy of Sciences, 
Warsaw 00-049, Poland \\
$^{b}$ Theoretisch-Physikalisches Institut  \\ 
\makebox[3mm]{ }Friedrich-Schiller-Universit\"at, 
Jena 07743, Germany   
\end{minipage} 

\vspace*{5mm} 
\centerline{\small \it 
(Submitted March 1999 --- 
Revised October 1999)}  


 \vspace*{-89mm}\vspace*{-2mm} 
\hbox to 6.25truein{
\footnotesize\it 
\hfil \hbox to 0 truecm{\hss 
\normalsize\rm 
{\sf  gr-qc/9912094} { }}\vspace*{-3.5mm}}
\hbox to 6.2truein{
\vspace*{-1mm}\footnotesize 
\hfil 
} 

  
\vspace*{65mm} \vspace*{16mm} 



\vspace*{0.5cm}

\begin{abstract}
\noindent
Recently proposed quantization in field 
theory based on an analogue of Hamilton\-ian formulation 
 which treats space and time on 
equal footing (the so-called De Donder-Weyl theory) 
is applied to General Relativity 
in metric variables.  
  We formulate a covariant 
analogue of the Schr\"odinger equation 
for the wave function of 
space-time and metric  variables 
and a supplementary ``bootstrap condition'' 
which 
 enables us to incorporate classical metric geometry 
as an approximate notion - a result of 
quantum averaging 
  - in 
the 
self-consistent with 
 the underlying quantum dynamics way.  
In this sense an independence of an arbitrarily chosen metric background 
is ensured. 

\end{abstract}


\newcommand{\beq}{\begin{equation}}
\newcommand{\eeq}{\end{equation}}
\newcommand{\beqa}{\begin{eqnarray}}
\newcommand{\eeqa}{\end{eqnarray}}
\newcommand{\nn}{\nonumber}

\newcommand{\half}{\frac{1}{2}}

\newcommand{\xt}{\tilde{X}}

\newcommand{\uind}[2]{^{#1_1 \, ... \, #1_{#2}} }
\newcommand{\lind}[2]{_{#1_1 \, ... \, #1_{#2}} }
\newcommand{\com}[2]{[#1,#2]_{-}} 
\newcommand{\acom}[2]{[#1,#2]_{+}} 
\newcommand{\compm}[2]{[#1,#2]_{\pm}}

\newcommand{\lie}[1]{\pounds_{#1}}
\newcommand{\co}{\circ}
\newcommand{\sgn}[1]{(-1)^{#1}}
\newcommand{\lbr}[2]{ [ \hspace*{-1.5pt} [ #1 , #2 ] \hspace*{-1.5pt} ] }
\newcommand{\lbrpm}[2]{ [ \hspace*{-1.5pt} [ #1 , #2 ] \hspace*{-1.5pt}
 ]_{\pm} }
\newcommand{\lbrp}[2]{ [ \hspace*{-1.5pt} [ #1 , #2 ] \hspace*{-1.5pt} ]_+ }
\newcommand{\lbrm}[2]{ [ \hspace*{-1.5pt} [ #1 , #2 ] \hspace*{-1.5pt} ]_- }
\newcommand{\pbr}[2]{ \{ \hspace*{-2.2pt} [ #1 , #2 ] \hspace*{-2.55pt} \} }
\newcommand{\we}{\wedge}
\newcommand{\dv}{d^V}
\newcommand{\nbrpq}[2]{\nbr{\xxi{#1}{1}}{\xxi{#2}{2}}}
\newcommand{\lieni}[2]{$\pounds$${}_{\stackrel{#1}{X}_{#2}}$  }

\newcommand{\rbox}[2]{\raisebox{#1}{#2}}
\newcommand{\xx}[1]{\raisebox{1pt}{$\stackrel{#1}{X}$}}
\newcommand{\xxi}[2]{\raisebox{1pt}{$\stackrel{#1}{X}$$_{#2}$}}
\newcommand{\ff}[1]{\raisebox{1pt}{$\stackrel{#1}{F}$}}
\newcommand{\dd}[1]{\raisebox{1pt}{$\stackrel{#1}{D}$}}
\newcommand{\nbr}[2]{{\bf[}#1 , #2{\bf ]}}
\newcommand{\der}{\partial}
\newcommand{\oo}{$\Omega$}
\newcommand{\Om}{\Omega}
\newcommand{\om}{\omega}
\newcommand{\eps}{\epsilon}
\newcommand{\si}{\sigma}
\newcommand{\Lm}{\bigwedge^*}

\newcommand{\inn}{\hspace*{2pt}\raisebox{-1pt}{\rule{6pt}{.3pt}\hspace*
{0pt}\rule{.3pt}{8pt}\hspace*{3pt}}}
\newcommand{\sro}{Schr\"{o}dinger\ }
\newcommand{\bm}{\boldmath}
\newcommand{\vol}{\omega}
               \newcommand{\dvol}[1]{\der_{#1}\inn \vol}

\newcommand{\bd}{\mbox{\bf d}}
\newcommand{\bder}{\mbox{\bm $\der$}}
\newcommand{\bI}{\mbox{\bm $I$}}

\newcommand{\be}{\beta} 
\newcommand{\ga}{\gamma} 
\newcommand{\de}{\delta} 
\newcommand{\Ga}{\Gamma} 
\newcommand{\gmu}{\gamma^\mu}
\newcommand{\gnu}{\gamma^\nu}
\newcommand{\ka}{\kappa}
\newcommand{\hka}{\hbar \kappa}
\newcommand{\al}{\alpha}
\newcommand{\lapl}{\bigtriangleup}
\newcommand{\psib}{\overline{\psi}}
\newcommand{\Psib}{\overline{\Psi}}
\newcommand{\derts}{\stackrel{\leftrightarrow}{\der}}
\newcommand{\what}[1]{\widehat{#1}}

\newcommand{\bx}{{\bf x}}
\newcommand{\bk}{{\bf k}}
\newcommand{\bq}{{\bf q}}

\newcommand{\omk}{\omega_{\bf k}} 

\newcommand{\lpl}{\ell}
\newcommand{\zb}{\overline{z}} 


\section{\hspace{-4mm}.\hspace{2mm}Introduction }

Among the conceptual problems of quantum gravity the 
so-called ``Issue of Time'' 
(see, e.g., \cite{isham})  
is perhaps the most quintessential 
one. 
 It originates in difficulties with reconciling 
 the basic principles of quantum theory with those of 
 General Relativity. One of these 
is the fact that, 
in contradistinction 
 with 
the space-time unity in relativity, 
in quantum theory the time dimension is singled out 
by the probabilistic interpretation of the formalism, 
the procedure of  canonical quantization, and 
the formulations of quantum evolution laws. 
 It is not unexpected then that the gravitation, 
by very its general relativistic nature,   
resists  to 
 getting quantized in the way discriminating  
between space and time dimensions. 
 Besides, 
the inherent nonlinearity of gravitation on the classical level 
makes the applicability of traditional pertubative techniques 
of quantum field theory rather doubtful, as it is also 
demonstrated  by the non-renormalizability problem.   
It would 
be more natural, therefore, to apply a nonpertubative 
quantization which 
 would treat 
space and time variables 
on equal footing. Thought this idea is partially incorporated 
in the path integral quantization, the latter essentially 
serves as  a basis of pertubative construction and still 
implicitly singles out the time parameter on the 
 level of interpretation, 
for example,  by referring to the {\em time-ordered }  
Green functions produced by the generating functional.

It is well known that the preferred  role of the time dimension 
appears 
already in the   classical Hamilton\-ian formalism which underlies  
canonical quantization. It leads to the picture of fields 
as infinite dimensional Hamiltonian systems evolving in time.  
 The approach of the present paper stems from the 
 (not yet commonly acknowledged in physics) fact  
that  
another generalization of the Hamiltonian 
formulation from mechanics to field theory is possible, 
 which treats space and time variables 
 on equal footing as analogues of the single time parameter 
in mechanics.  
The simplest realization of this 
idea is the so-called De Donder-Weyl (DW) 
 formulation 
of the field equations \cite{dw}: 
given  a Lagrangian density 
$L=L(y^a, y^a_\mu, x^\nu)$, a function of 
field variables $y^a$, their space-time derivatives 
(first jets) $y^a_\mu$ and   space-time 
variables $x^\mu$,  
we can introduce new Hamiltonian-like variables  
\beq 
p_a^\mu:=\der L / \der (\der_\mu y^a) \quad \mbox{\rm and} \quad   
H=H(y^a,p^\mu_a,x^\nu):=\der_\mu y^a p^\mu_a -L,   
\eeq  
to be referred to as {\em polymomenta} 
and the {\em DW Hamiltonian function}, respectively, 
and write the 
field equations 
in the DW canonical form 
\beq
\der_\mu y^a  = \frac{\der H}{\der p^\mu_a}, 
\quad 
\der_\mu  p^\mu_a=- \frac{\der H}{\der y^a }  . 
\eeq

In formulation (1.2) fields are treated as ``multi-time'' 
generalized (=DW) Hamilton\-ian systems {\em varying} in 
space-and-time (rather than evolving in time).  
No distinction between space and time dimensions,  
i.e. no topological restriction to globally hyperbolic space-times,  
is implied.  
Moreover, another potential advantage of 
the formulation (1.2)  is that it manages to describe  
the dynamics of a field, usually 
 viewed as  
an infinite dimensional 
mechanical system, 
in a finite dimensional analogue of the phase space, 
the space of variables $(y^a, p_a^\mu, x^\nu)$. This, of course, does not 
mean a loss in the number of degrees of freedom: just instead of 
the  notion of a degree of freedom per space point, 
which is inspired by the conventional Hamiltonian treatment, 
 here the number of components of the field 
over space-time 
is more relevant. The equivalence of (1.2) 
to the Euler-Lagrange equations, which is only restricted 
by the regularity of DW Legendre transform: 
$y^a_\nu \rightarrow p_a^\nu, 
L \rightarrow H$,  
proves {\em de facto} that no degrees of freedom get lost.

It is natural to inquire whether 
the above DW Hamiltonian formulation can be 
a basis of quantization of field theory in general 
and  of General Relativity in particular.

The structures  of  canonical formalism which are   
used in various 
quantization procedures 
 have been 
generalized to DW formulation 
of field theory in  
\cite{romp98,bial96}. This generalization  
leads to Poisson brackets defined on differential forms playing 
the role of dynamical variables and to a graded analogue of a 
Poisson algebra: the so-called Gerstenhaber algebra and its 
generalizations. 
 Elements of the corresponding field quantization 
have been 
considered in \cite{qs96,bial97,lodz98}. 
In short, in the 
$y$-representation 
one is lead to the realization of polymomenta as operators 
\beq
p_a^\mu = -i\hbar\kappa \gamma^\mu \frac{\der}{\der y^a} , 
\eeq 
which act on spinor (or the space-time Clifford algebra valued) 
wave functions on the configuration space of field and 
space-time  variables: $\Psi=\Psi(y^a,x^\mu)$; 
$\gamma^\mu${}'s  denote the imaginary units of the space-time 
Clifford algebra.  
The constant 
$\kappa$ of the dimension [length]$^{-(n-1)}$ ensures the dimensional 
consistency of (1.3) and is interpreted as a quantity of the ultra-violet 
cutoff or the fundamental length scale \cite{qs96,lodz98}. 
Note that in this picture all the space-time dependence in 
transfered from operators to the wave function corresponding 
to what could be called the ``ultra-Schr\"odinger'' picture. 
The dynamical law for the wave function was proposed to have 
the form of a generalized covariant Schr\"odinger equation 
\beq
i\hbar\kappa \gamma^\mu \der_\mu \Psi = \what{H}\Psi 
\eeq
where $\what{H}$ is the operator form of the DW Hamiltonian function. 
For example, for scalar fields $y^a$ with 
$L= \frac{1}{2} \der_\mu y^a \der^\mu y^a - V(y)$  
we obtain $H= \frac{1}{2}  p^\mu_a p_\mu^a + V(y)$, where 
$p_a^\mu=\eta^{\mu\nu}\der_\nu y^a$, 
and then 
$\what{H}= -\frac{1}{2}\hbar^2\kappa^2\der_a\der_a 
+ V(y)$. 
In \cite{bial97,lodz98} we argued that the above generalization 
of the Schr\"odinger equation fulfills some natural requirements 
 of 
the correspondence principle.  
In particular, it leads to an analogue of the Ehrenfest theorem  
and can be reduced 
to the field theoretic  DW Hamilton-Jacobi equation 
(with some additional conditions)  
 in the classical limit.   
It should be noticed, however, that in spite of this formal success  
the description of quantized fields 
provided by the present framework is very different from that 
in the conventional quantum field theory and its physical 
 significance remains to be investigated.  

The purpose of this paper is to 
 discuss quantization of General Relativity from the point of 
view of the above approach 
(see  \cite{jena99-1} for a somewhat more detailed discussion). 

\section{\hspace{-4mm}.\hspace{2mm} Quantizing General Relativity } 


{\bf 2.1 Extension to curved space-time 
and application to General Relativity. } 
The application to General Relativity  requires an extension 
 of the approach to curved space time  with the metric 
$g_{\mu\nu}(x)$. 
We introduce 
$x$-dependent  
 curved space-time 
Dirac matrices 
$\gamma_\mu(x)$ which fulfill 
$ 
\gamma_\mu(x) \gamma_\nu(x) + \gamma_\nu(x) \gamma_\mu(x)
=2 g_{\mu\nu}(x)      
$ 
and can be expressed in terms  of the 
Minkowski space Dirac matrices 
$\gamma^A\!: \, \gamma^A\gamma^B + \gamma^B\gamma^A 
= 2 \eta^{AB}, $ 
and the vielbein coefficients $e_A^\mu (x)\!: \, 
e_\mu^A (x) e_\nu^B (x) \eta_{AB}:=g_{\mu\nu}(x) $. 
Then the curved space-time version of 
(1.4) assumes the form 
\beq  
i\hbar\kappa 
\gamma^\mu (x) \nabla_\mu \Psi 
= \what{H} \Psi 
\eeq
where 
$\what{H} $ is 
the 
operator of DW Hamiltonian 
function      
and 
$\nabla_\mu$ is the covariant derivative.  
If $\Psi$ is a spinor wave function 
$\nabla_\mu =  \der_\mu + \theta_\mu$ 
 is the spinor 
covariant derivative with the spinor connection 
$ 
\theta_\mu = \frac{1}{4} \theta_{AB}{}_\mu \gamma^{AB},
$ 
where 
$\gamma^{AB}:= \half (\gamma^A\gamma^B - \gamma^B\gamma^A)$ 
and 
\beq
\theta^A{}_{B\mu} = 
 e^A_\al e^\nu_B \Ga^\al{}_{\mu\nu} 
 -e^\nu_B \der_\mu e^A_\nu .
\eeq 

In General Relativity  treated  as the field theory of  
 the metric field   
the DW configuration space 
 is 
the   bundle of symmetric rank-two 
(metric) tensors $g^{\mu\nu}$ 
over 
space-time and, respectively, the wave function 
 is a function on this space. With 
field variables taken to be 
the components of the metric  the polymomenta will be some combinations 
of the connection 
coefficients and, therefore, 
according to (1.3), 
the corresponding operators  will involve 
differentiations with respect to the metric variables. Hence, a 
generalized Schr\"odinger equation for gravity 
 will assume the symbolic form (c.f. (2.1))  
\beq
i\hbar\kappa 
\what{ \mbox{$\hspace*{0.0em}e\hspace*{-0.3em}
\not\mbox{\hspace*{-0.2em}$\nabla$}$}} 
 \Psi = 
\what{{\cal H}\hspace*{-0.0em}} \Psi  ,   
\eeq 
 where   
$\what{{\cal H}}$ is the operator form 
of 
 DW Hamiltonian density  
of gravity, $\what{{\cal H}}:= \what{e H}$, 
$e:= |{\rm det}(e^A_\mu)|, $ 
and 
$\what{\mbox{$\hspace*{0.0em}\not\mbox{\hspace*{-0.2em}$\nabla$}$}}$  
denotes the quantized Dirac operator in the sense that 
the corresponding connection coefficients are 
replaced by appropriate differential operators.   
A more specific form of this equation will be derived below. 
  Note that quantum gravity 
 possesses an intrinsic fundamental length scale, the Planck length 
$\lpl$, 
 suggesting 
an  identification of 
the parameter $\kappa$ 
 with the Planck scale quantity: 
 $\kappa \sim \lpl^{-(n-1)}$.  

It should be noticed  
 that if the wave function in (2.3) is spinor valued 
the Dirac operator involves the  spinor connection 
which classically depends on the first derivatives of vielbeins 
(c.f. eq. (2.2)).  
To quantize this object we need to Legendre transform 
first derivatives of vielbeins 
to corresponding polymomenta and then quantize the latter. 
This procedure is only possible within the vielbein formulation of General 
Relativity, which would be, therefore, 
the most suitable framework for discussing the 
application of the present approach  
to gravity. However, so far 
there is no appropriate DW-like formulation of General Relativity 
in vielbein variables 
  in our disposal. 
For this reason 
 the following discussion,  as a first step,   
is based on the metric formulation, 
which, nevertheless, also appears to be useful.  

%

\noindent 

{\bf 2.2 DW-like formulation of General Relativity. } 
The most straightforward way to represent the Einstein equations 
in DW form (1.1) is to start from their first 
order form: symbolically,   
$\der \Gamma + \Gamma = 0$, { } { }
$\der g + \Gamma = 0$,         
where $\Gamma$'s denote different appropriate 
linear combinations of the  Christoffel symbols and 
$g$ denotes a  combination of  metric variables,  
and to find the 
combinations such  that 
 the Einstein equations assume the  form 
\beq  
\der_\al h^{\be\ga}
= 
\der {\cal H}  / \der Q^\al_{\be\ga} ,  
\quad 
\der_\al Q^\al_{\be\ga}
=  
- \der {\cal H}   / \der h^{\be\ga}  .   
\eeq   
 In this formulation 
 the 
field variables 
  are 
$ h^{\alpha\beta}:= \sqrt{g}g^{\al\be}$, 
where $g:={|\rm det}(g_{\mu\nu})|$, 
the polymomenta 
  are 
\beq 
Q^\al_{\beta\gamma} := 
 \frac{1}{8\pi G}( \delta^\alpha_{(\beta}\Gamma^\delta_{\gamma)\delta} 
  - \Gamma^\al_{\beta\gamma}) ,    
\eeq  
and the DW Hamiltonian density 
  takes the form 
\beq
{\cal H} 
(h^{\alpha\be}, Q^\al_{\beta\gamma}) :=  
 8\pi G\, 
h^{\alpha\ga} \left ( 
Q^\de_{\al\be } Q^\be_{\ga\de }+ 
\frac{1}{1-n}\, Q^\be_{\al\be }Q^\de_{\ga\de } \right )
+(n-2) \Lambda \sqrt{g} .   
\eeq

 Note that 
the above formulation of the Einstein equations has a deeper foundation 
in the theory of Lepagean equivalents in the calculus of variations 
\cite{horava}. 

\noindent 

{\bf 2.3 Local quantization. }  
By formally following the curved space-time version of our 
scheme  we obtain the operator form of polymomenta  
\beq
\what{Q}{}^\al_{\be\ga} = -i\hbar \kappa 
\gamma^\alpha  
\left \{ \sqrt{g} 
\frac{\der}{\der h^{\beta\gamma}} \right \}_{ord}  
\eeq 
which is determined up to an 
ordering of the expression inside 
the curly brackets. 
Then, plugging (2.7) into (2.6) and assuming 
in the intermediate calculation
the standard ordering of operators 
 we obtain 
 the operator form of the DW Hamiltonian 
density  
\beq
\what{{\cal H}} 
= - 8\pi G\, 
\hbar^2\kappa^2 \frac{n-2}{n-1}  
\left \{  
 \sqrt{g} 
h^{\al\ga}h^{\be\de}\frac{\der}{\der h^{\al\be}} 
\frac{\der}{\der h^{\ga\de}} \right \}_{ord} 
+ (n-2)  \Lambda 
 \sqrt{g} 
\eeq  
which is also ordering dependent. 

In should be noted, however, 
that  the right hand sides of Eqs. (2.7) and (2.8) are tensorial 
while the left hand sides are not: 
polymomenta $Q{}^\al_{\be\ga}$  transform as 
the connection coefficients and ${\cal H}$, 
being 
 essentially the truncated Einstein-Hilbert Lagrangian density,  
is not a diffeomorphism scalar. 
Therefore, 
operator realizations (2.7) and (2.8) can be valid only 
locally in a specific 
coordinate system. 
This is, however, in accordance 
with the fact that the Poisson brackets underlying the 
rule of quantization (1.3) (see, e.g.,  \cite{bial97}) 
can be interpreted as, in a sense, ``equal-point''. 
 This is 
the Schr\"odinger equation (2.3) 
 which 
is supposed to tell us how to go from one point 
of 
space-time to another. 
 This equation, 
however,  contains the 
spinor connection coefficients which classically have the form (2.2). They 
obviously cannot be fully quantized within the metric formulation for, 
the second term in (2.2) cannot be expressed in terms of the polymomenta 
of the metric formulation. 

To quantize the connection coefficients let us note that classically 
(c.f. (2.5)) 
\beq 
\Gamma^\al_{\be\ga}= 8\pi G 
\left ( 
\frac{2}{n-1} 
\delta^\al_{(\beta } Q^\delta_{\ga)\delta} 
- Q^\al_{\be\ga} 
\right )  . 
\eeq 
Using (2.7) we can write for the operator $\what{\Gamma}{}^\al_{\be\ga}$   
an ordering dependent expression 
\beq  
\what{\Gamma}{}^\al_{\be\ga} = 
- 8\pi i G\hbar\kappa \sqrt{g} 
\left 
(\frac{2}{n-1} \delta^\al_{(\beta } \gamma^\sigma 
\frac{\der}{\der h^{\ga ) \sigma }} 
- \ga^\al \frac{\der}{\der h^{\beta \ga}} 
\right ) 
+ \tilde{\Gamma}{}^\al_{\be\ga}(x) 
\eeq 
where the auxiliary connection 
$\tilde{\Gamma}{}^\al_{\be\ga}(x)$ is introduced to make the 
operator $\what{\Gamma}{}^\al_{\be\ga}$ transforming like 
a connection. We see now that (2.7) and (2.8) can be 
viewed 
as valid locally (in the vicinity of a point $x$) 
in the geodesic coordinate system in which 
$\tilde{\Gamma}{}^\al_{\be\ga}|_x = 0$. In the same coordinate 
system the locally valid   Schr\"odinger equation 
can be written in the 
form (c.f. (2.3)) 
\beq 
i\hbar\kappa \sqrt{g}\gamma^\mu 
(\der_\mu + \hat{\theta}_\mu)  
\Psi = 
\what{{\cal H}} \Psi 
\eeq 
where  the operator form of the spinor connection coefficients 
is given by (c.f. (2.2)) 
\beqa 
\what{\theta}^A{}_{B\mu} &=&  
-8\pi i G \hbar \kappa \sqrt{g} 
e^A_\al e^\nu_B 
\left ( 
\frac{2}{n-1} \delta^\al_{(\mu } 
\gamma^\sigma 
\frac{\der}{\der h^{\nu)\sigma }} 
- \ga^\al \frac{\der}{\der h^{\mu \nu}} 
\right ) 
+\tilde{\theta}^A{}_{B\mu}|_x  \nn \\
&=:& ({\theta}^A{}_{B\mu})^{op} + 
\tilde{\theta}^A{}_{B\mu}|_x 
\eeqa  
and involves the ordering dependent operator part 
$({\theta}^A{}_{B\mu})^{op}$  
and a non-vanishing  (even though 
$\tilde{\Gamma}^\al_{\be\ga}|_x =0$) 
auxiliary spinor connection part 
$\tilde{\theta}^A{}_{B\mu}|_x$. 

\noindent 
{\bf 2.4 Covariant Schr\"odinger equation and the 
``bootstrap condition.'' } 
To formulate a diffeomorphism covariant version of (2.11) 
we notice 
that vielbeins do not 
enter the present DW Hamiltonian 
formulation of General Relativity 
and, therefore, 
within the present consideration 
may (and can only) be 
treated 
as non-quantized classical 
$x$-dependent  
 quantities.  
The  reference vielbein field,  
$\tilde{e}{}^\mu_A(x)$, 
however, is not quite arbitrary. The correspondence principle 
requires it to be consistent with the mean value of the metric  
tensor in the sense that 
\beq  
\tilde{e}{}^\mu_A(x)
\tilde{e}{}^\nu_B(x) \eta^{AB} = 
\left < g^{\mu\nu}\right >(x) .    
\eeq   
The latter 
is given by 
averaging  over the space of 
 the 
metric components by means of 
the wave function $\Psi(g^{\mu\nu},x^\mu)$:   
\beq
\left <g^{\mu\nu}\right >(x) = 
\int [d g^{\al\be}] 
~\Psib (g,x) g^{\mu\nu} \Psi(g,x)  ,   
\eeq 
where 
the invariant integration measure 
in  $\frac{1}{2}n(n+1)$--dimensional space of metric components 
  reads (c.f. \cite{misner}) 
\beq
[d g^{\al\be}] = 
\sqrt{g}^{\,(n+1)} \prod_{\al \leq \be} d g^{\al\be} . 
\eeq   

Therefore, 
the metric 
geometry explicitly appears as 
 a 
result of  
quantum averaging of 
the metric operator $g^{\mu\nu}$.      
 The local orientation of 
vielbeins  remains unquantized 
 and is supposed 
to be exclusively due to a choice of 
 a 
reference 
vielbein field 
(local reference frames)   
 which, however,  
 is restricted by the consistency with the 
averaged metric 
 according to 
the ``bootstrap condition'' (2.13).

Now, 
a generally covariant version of (2.11) takes  the form 
\beq
i\hbar\kappa 
\tilde{e} 
\tilde{e}^\mu_A(x)\ga^A(\der_\mu + 
\tilde{\theta}_\mu (x)) \Psi  
+ 		
i\hbar\kappa (\sqrt{g} \gamma^\mu {\theta_\mu})^{op} \Psi 
=  \what{{\cal H}}\Psi  , 
\eeq
where the term related to the quantized part 
$\theta_\mu^{op}$ of the total 
spinor connection   
$\theta_\mu = \tilde{\theta}_\mu (x) + \theta_\mu^{op}$ 
has, up to an ordering of operators,   
the form 
\beq 
(\sqrt{g} \gamma^\mu {\theta_\mu})^{op} 
= - n 
\pi i G \hbar\kappa 
\left \{ \sqrt{g} h^{\mu\nu} 
\frac{\der}{\der  h^{\mu\nu}} \right \}_{ord}  
\eeq  
 and the $x$-dependent reference spinor connection term 
$\tilde{\theta}_\mu (x)$ 
in (2.12) can be calculated using 
the reference vielbein field 
$\tilde{e}{}^A_\mu (x)$ consistent with (2.13)   
and  the 
classical expression 
\beq
\tilde{\theta}{}_\mu^{AB} (x) =  
\tilde{e}{}^{\al [A} 
\left 
( 
 2 \der_{[\mu} \tilde{e}{}^{B ]}{}_{\al ]} 
+ \tilde{e}{}^{ B] \be} \tilde{e}{}^C_\mu \der_\be 
\tilde{e}{}_{C \al}  
\right )
\eeq
which is equivalent to (2.2). 

To complete the description, 
a gauge-type 
condition 
 should be imposed in order to distinguish the physically relevant 
information. For example,  
the De Donder-Fock harmonic gauge   
\beq 
\der_\mu \left <\sqrt{g}g^{\mu\nu}\right >(x) = 0  
\eeq 
can be chosen. 
 Notice, that this is  
a gauge condition on the wave function 
$\Psi(g^{\mu\nu}, x^\nu)$ rather than on 
the metric field.

Let us note that eq. (2.16) can be written in 
a more compact form 
\beq
i\hbar\kappa 
\widetilde{ \mbox{$\hspace*{0.0em}e\hspace*{-0.3em}
\not\mbox{\hspace*{-0.2em}$\nabla$}$} 
} 
\Psi 
+  
i\hbar\kappa (\sqrt{g}  \gamma^\mu {\theta_\mu})^{op} \Psi 
=  \what{{\cal H}}\Psi   ,   
\eeq  
where $
\widetilde{ \mbox{$\hspace*{0.0em}e\hspace*{-0.3em}
\not\mbox{\hspace*{-0.2em}$\nabla$}$}} 
$ denotes the ``averaged'' Dirac operator (multiplied by the density $e$)   
in which the vielbein 
fields and the corresponding spinor connection are 
assumed to be consistent 
with the ``bootstrap condition'' (2.13).  
Obviously, the resulting equation for quantized gravity is 
integro-differential and nonlinear in essence.   
 The meaning of this description is that 
the metric of the space-time on which the wave function propagates 
is self-consistent with the quantum dynamics of the latter.  
The space-time metric geometry arises as an approximate 
notion -- a result of quantum averaging -- and is used as such 
in the left hand side of (2.20) 
 to describe the wave function propagation. 
 In this sense the formulation is independent of an 
 arbitrarily chosen metric background. 
 At the same time,    
 no 
explicit description of what could be thought to be an 
underlying   ``quantum pre-geometry'' 
 has been used. 
 A possible speculation could be 
that the 
appearance of the averaged Dirac operator in the left hand side of 
(2.20) 
may imply an approximate, 
not ultimately quantum, character of the description achieved 
here, which may neccessiate 
a further step in which the space-time itself, 
not only the metric structure, 
would be  treated quantum theoretically.  
 This is, in fact, a rather common expectation in quantum gravity. 
 Our eq. (2.20) implies that the idea requires a generalized notion 
 of the Dirac operator beyond the framework of classical manifolds. 
For example, non-commutative geometry \cite{connes} 
provides a framework for treating the 
Dirac operator in the left hand side of (2.3) or (2.20) 
and the underlying geometry 
in a ``non-commutative'' fashion, very much in the spirit 
of quantum theory. String/M- theory also 
enables us to consider  
quantized space-time coordinates 
(see, e.g., \cite{witten}).  
An analysis of these perspectives could be a subject 
of a subsequent research. 

\section{Conclusion }

We have argued 
that within the quantization scheme based on DW theory 
quantum General Relativity 
may be described by Eq. (2.16), 
with  operators $\what{{\cal H}}$ and 
$(\sqrt{g} \gamma^\mu {\theta_\mu})^{op} $ given,  
respectively, by 
Eq. (2.8) and Eq. (2.17),    
and 
the 
bootstrap condition (2.13) 
ensuring self-consistency and, effectively, background independence.   
These equations 
 define 
the wave function $\Psi(g^{\mu\nu}, x^\al)$ 
which  
may be interpreted as the probability 
amplitude to find the values of the components 
of the metric in the interval 
$[g^{\mu\nu}$ -- $(g^{\mu\nu} \!+\! dg^{\mu\nu}) ]$ 
in an infinitesimal vicinity of the point $x^\al$. 
Obviously, this description is very different from 
the conventional quantum field theoretic one and 
its physical  
 implications 
remain to be  analysed. 
Note, however, that it opens an intriguing 
possibility to 
approximate  
the ``wave function of the Universe'' 
by a 
solution of  equation (2.16) 
which expands from  an initially localized wave packet   
making the observation of 
space-time points beyond the primary  ``probability clot'' 
more and more probable, in this sense attributing a meaning to the 
process of a ``genesis of space-time''. 
 Obviously, an essential role in this process  
is played by 
the self-consistency imposed by the 
bootstrap condition (2.13).

\bigskip

\noindent 
{\bf Acknowledgments }
\medskip 

I thank Prof. A. Borowiec and Prof. J. Klauder 
for their valuable remarks at the Bia\l owie\.za'98 Workshop.  
I gratefully acknowledge the 
Institute of Theoretical Physics of the 
 Friedrich Schiller 
University (Jena, Germany)     
and Prof. A. Wipf   
\, for kind hospitality and excellent 
working conditions which   
only enabled me to finish this paper.  
   
\noindent  


\end{document}